\documentclass[sn-mathphys-num]{sn-jnl}

\usepackage{graphicx}%
\usepackage{multirow}%
\usepackage{amsmath,amssymb,amsfonts}%
\usepackage{amsthm}%
\usepackage{xcolor}%
\usepackage{textcomp}%
\usepackage{manyfoot}%
\usepackage{booktabs}%
\usepackage{algorithm}%
\usepackage{afterpage}
\usepackage{algorithmicx}%
\usepackage{algpseudocode}%
\usepackage{listings}%
\usepackage{tikz}
\usepackage{tabularx}
\usetikzlibrary{shapes.geometric, arrows, positioning}
\usepackage{array, makecell} 
\usepackage{soul}
\usepackage{spverbatim}
\usepackage{pgfplots}
\pgfplotsset{width=10cm,compat=1.16}
\usepackage{placeins}

\begin{document}

\title[A Comprehensive Survey and Classification of Evaluation Criteria for Trustworthy Artificial Intelligence]{A Comprehensive Survey and Classification of Evaluation Criteria for Trustworthy Artificial Intelligence}

\author*[1]{\fnm{Louise} \sur{McCormack}}\email{louise.mccormack@adaptcentre.ie}

\author[1]{\fnm{Malika} \sur{Bendechache}}\email{Malika.bendechache@adaptcentre.ie}
\affil[1]{\orgdiv{ADAPT Research Centre}, \orgname{School of Computer Science, University of Galway}, \orgaddress{\city{Galway}, \country{Ireland.}}
\\
\\\textbf{Note:} This work has been accepted for publication in AI and Ethics}


\abstract{This paper presents a systematic review of the literature on evaluation criteria for Trustworthy Artificial Intelligence (TAI), with a focus on the seven EU principles of TAI. This systematic literature review identifies and analyses current evaluation criteria, maps them to the EU TAI principles and proposes a new classification system for each principle. The findings reveal both a need for and significant barriers to standardising criteria for TAI evaluation. The proposed classification contributes to the development, selection and standardization of evaluation criteria for TAI governance.}


\maketitle
\section{Introduction}\label{sec1}
AI has become prominently used for decision-making across various sectors, including finance, healthcare, judiciary, and insurance\cite{Paganoetal_2023}. 72\% of businesses are now using AI in at least one business function, with a spike in generative AI adoption which doubled in ten months to a 65\% adoption rate\cite{McKinsey2024}. Additionally, black box algorithms are increasingly being commercially employed to make critical safety decisions in areas such as self-driving cars and healthcare, leaving us at risk of relying on systems that we do not fully understand\cite{Guidottietal_2018}. Algorithms also personalize news content, user-generated content, and advertisements for individuals through tailored social media feeds, search engines and news platforms, calling attention to algorithmic accountability for content personalization\cite{Descampeetal_2022}. Consumers have free access to powerful, user-friendly AI tools using conversational interfaces resembling human interactions, such as tools powered by OpenAI's ChatGPT large language model (LLM). 

Despite the expanding sophistication and impact of AI systems across different sectors, there are concerns about their ethics, safety, and trustworthiness. A survey by the Pew Research Centre\cite{smithpew_2018} shows that about 58\% of the public expresses broad concerns about the fairness and acceptability of using AI for decision-making in situations with critical real-world consequences. About 40\% of those surveyed also believe that AI decision-making could be developed bias-freely.  

Several initiatives are established to address incorporating ethics into TAI, focusing on developing frameworks and guidelines for the governance of AI systems\cite{PiersonHildt2023}. These frameworks acknowledge a need for ethical considerations in designing and operating advanced systems. Notable contributions are the European Commission's Trustworthy AI Ethics Guidelines\cite{AIHLEG2019} and its subsequent Assessment List for Trustworthy Artificial Intelligence (ALTAI)\cite{EUALTAI_2020}, which organizes foundational ethical considerations for AI\cite{Golpayeganietal_2023}. TAI is now starting to impact business; for example, the Deloitte Trustworthy AI Framework is being used to implement ethical principles to AI projects\cite{Deloitte_2020}. Despite these early steps, there is still a clear need for a standardized evaluation criteria AI for trustworthiness\cite{Kozuka_2019}\cite{deAlmeidaetal_2021}\cite{Schneideretal_2023}. In December 2023, the International Organization for Standardization published an international standard, ISO/IEC 42001\cite{ISO42001}, which offers a high-level specification for Artificial Intelligence Management Systems. This ISO is considered an essential document for TAI with the potential to assist conformity to the EU AI Act. 

Several surveys returned in our search query discussed the evaluation of individual TAI principles; however, these papers generally use surveying techniques that improve or offer classifications for these TAI principles. For example, Ali et al.\cite{Alietal_2023} surveyed the area of Explainable AI (XAI) and proposed both an XAI classification and an evaluation criterion to assess XAI within AI systems. Cooper et al.\cite{Cooperetal_2022} used the existing research into Nissenbaum's "four barriers" to accountability to frame an assessment approach for accountability and fairness in machine learning (ML). While the authors did not provide an assessment criterion to evaluate or score accountability, their paper did give a set of conditions required for a moral and relational accountability framework for AI systems, which is a starting point for assessing these systems for accountability. Ma et al. (2022)\cite{Maetal_2022} published a survey on privacy and security for distributed learning in ML. The authors provided a comprehensive overview of the state-of-the-art attack mechanisms and defensive techniques, including multiple quantifiable evaluation techniques for privacy and security in AI. Zhou et al.\cite{Zhouetal_2022} and Pagano et al.\cite{Paganoetal_2023} published survey papers on fairness metrics and methods for mitigating bias. Ma et al.\cite{Maetal_2022} published a survey on privacy and security, which reviews the state-of-the-art attack mechanisms and defensive techniques for ML. These papers consider many evaluation criteria, including quantifiable metrics for individual TAI principles, such as fairness or privacy. Although these surveys discuss approaches for assessment and evaluation criteria, none of these papers are primarily focused on scoring and evaluating their specific Trustworthy AI Principle. Each of the referenced survey papers focuses mainly on offering a classification or evaluating methods and techniques for mitigating risks rather than focusing on the evaluation criteria.

The diversity, non-discrimination, and fairness surveys were the most advanced in providing quantifiable evaluation criteria. These three terms are used interchangeably in the literature, referring to the same thing\cite{PessachShmueli_2022}. This paper will use fairness as a general classification for this area, with a definition provided in table\ref{tab2}. For further reading, a comprehensive classification of types of bias and comparisons of fairness definitions can be found in in the paper published by Mehrabi et al.\cite{Mehrabietal_2021}. Outside of fairness, we found a lack of surveys offering developed evaluation criteria for other TAI principles. However, two surveys considered Trustworthy AI through all seven principles.
Liu et al.\cite{Liuetal_2022} conducted a comprehensive appraisal of Trustworthy AI from a computational perspective considering all seven EU principles. They compiled an extensive section on concepts and taxonomy for six areas and outlined the relevant technical methods. They proposed classifying TAI principles based on technical, user or social perspectives. This survey offers a good starting point for areas for assessment, particularly with the inclusion of some real-world applications. However, they do not focus on evaluation criteria for TAI principles.
 
The other survey paper by Chamola et al.\cite{Chamolaetal_2023} is a survey about technologies and current use cases of AI through the lens of the seven EU principles on Trustworthy AI. While this survey provides an excellent summary of available technology in these areas, particularly in XAI, it does not focus on evaluation criteria. Indeed, the authors' concluding remarks note finding an absence in the literature of any specified criteria for assessing trustworthy AI. 
 
Our paper provides a systematic literature review of existing metrics for evaluating and scoring AI systems for trustworthiness. We summarize the latest research in each area using the lens of the seven EU TAI principles. For fairness, we provide a classification of evaluation criteria, including a range of established fairness metrics. For the remaining six less-researched TAI principles, we provide starting points for evaluating them by including the latest research relating to the review of these principles, noting any metrics where available in the literature. In addition, we discuss the assessment of multiple TAI principles, including a discussion of trade-offs between principles. 
 
The rest of the paper is organized as follows. Section\ref{sec2} provides a background to the area of assessment of Trustworthy AI. Section \ref{sec3} details our methodology and research questions. Section \ref{sec4} covers the findings of this paper, including the proposed Trustworthy AI evaluation criteria classification and detailed analysis. The classification consists of a breakdown of all the metrics in the literature to evaluate the seven TAI principles. Section \ref{sec5}, outlines the issues and challenges for TAI evaluation. In section \ref{sec6}, we provide our conclusion and recommendations for future research.

\section{Background}
\label{sec2}
The following section discusses the fundamental concepts of Trustworthy AI (TAI) and provides a background to TAI assessment. 

TAI refers to ethical considerations that must be considered when building and using AI systems\cite{Kauretal_2022}. AI4People published their core principles for AI in their Ethical Framework for a Good AI Society in 2018\cite{Floridietal_2018}, foundational for the subsequent work undertaken in this area, including the work by the EU. Their five critical metrics for trustworthy AI were fairness and non-discrimination, technical robustness and safety, transparency and explainability, accountability and human oversight. They offered vital advice for trustworthy AI, including what they counselled as the most crucial recommendation, which was to design an explicit process for assessing AI risk and mitigating it for each application of AI, explicitly calling out the need for the development of agreed-upon Trustworthy AI metrics to enable user-driven benchmarking. Some participants later challenged the success of the AI4People programme, highlighting a lack of diversity of experts involved in the media industry\cite{PiersonHildt2023}. The European Commission's independent High-Level Expert Group for Artificial Intelligence (AI HLEG) published Ethics Guidelines for Trustworthy AI\cite{AIHLEG2019} in 2019, which included seven principles that AI systems should adhere to. In July 2020, following a piloting scheme, the AI HLEG subsequently developed The Assessment List for Trustworthy Artificial Intelligence (ALTAI)\cite{EUALTAI_2020}, an assessment booklet and some proposed questions that aim to assist in actioning the seven TAI principles. It advised that TAI guidelines be implemented in all stages of the AI system. The seven TAI principles are human agency and oversight, technical robustness and safety, privacy and data governance, transparency, diversity, non-discrimination and fairness, societal and environmental well-being, and accountability\cite{EUALTAI_2020}. Building on this work, the EU drafted a proposal for the AI act. It was partly criticized because it did not provide an assessment criterion and left businesses responsible for self-assessment\cite{PiersonHildt2023}. ISO/IEC 42001\cite{ISO42001} offers a high-level specification for Artificial Intelligence Management Systems and is considered a key document to consider concerning conformity with the EU AI Act\cite{Golpayeganietal_2023}. 

There is a clear need for a governance framework to measure trustworthy AI\cite{Kozuka_2019}\cite{deAlmeidaetal_2021}\cite{Schneideretal_2023}. Haupt et al. (2021) \cite{Hauptetal_2021} called for more standardized metrics for trustworthy AI, noting that while some standards existed for evaluating various AI trust metrics in their use case of ML in weather and climate, there was a lack of standardization around the metrics used. Standard identification and management is also outlined as a key step in the AI tool-supported audit process outlined by Ojewale et al.\cite{ojewale2024towards}. Their research into AI Audit Tooling also raised the need for metrics and benchmarks for AI evaluation.
 
We found that the high-level TAI governance frameworks we have outlined in this section were either influenced by or heavily overlapped with the seven EU TAI principles. For this reason, we chose to use these seven principles as a classification for summarising and commenting on assessment approaches and evaluation criteria for TAI.
 
When we refer to Trustworthy AI (TAI) throughout this paper, we do this through the lens of these seven EU principles. Trust metrics are metrics used to score AI within these seven TAI principles.

\subsection{Seven EU Principles for Trustworthy AI}\label{subsec1}
This section gives an overview of the seven ethical principles for developing and using AI systems proposed by the High-Level Expert Group informing European strategy on AI 2019\cite{AIHLEG2019}, which were developed further in 2020 in the assessment list for trustworthy AI (ALTAI)\cite{EUALTAI_2020}, and subsequently incorporated into the design of the EU AI Act\cite{eu2024aiact}. We also include a table of specific terms and outline the specific context for those terms as we refer to them throughout this paper.
\begin{itemize}
\item \textbf{Human agency and oversight:}
The first of the seven ethical AI principles is human agency and oversight. AI systems should be designed not to replace human autonomy but to enhance and support human decision-making. It advises that humans should be able to intervene in AI decisions when required. 

\item \textbf{Technical robustness and safety:}
AI systems should be dependable. This TAI principle stipulates that from a security perspective, they should be resilient to attack.  They should be designed to consider possible threats, and risks should be defined and mitigated as much as possible. The data used to train the AI should be accurate. There should be fallback plans in place. In cases where the AI is designed for continuous self-learning, fallback plans and constant security, robustness and safety reviews are essential to consider. 

\item \textbf{Privacy and data governance:}
This TAI principle considers the area of privacy and data governance. AI systems should be in line with privacy and data protection laws. The most relevant of these in the EU is the General Data Protection Regulation (GDPR). Where data is collected, it should be transparent about which data they collect, how it is used, and who can access that data.

\item \textbf{Transparency:}
Transparency, or explainable AI (XAI), is a crucial principle of trustworthy AI. AI systems should be explainable and understandable. Users should be able to understand the actions or decisions made by the AI quickly. Decision-making processes by the AI should be documented to allow for traceability. 

\item \textbf{Diversity, Non-discrimination, and Fairness:}
This TAI principle advises that training and decision-making should consider diversity, non-discrimination, and fairness. AI systems should be designed to avoid bias and discrimination in the decision design and how the AI gets trained. The system should also be designed to be inclusive and accessible to all, regardless of any personal protected classes, such as race or gender. They should incorporate elements of accessibility and universal design. Systems should be tested for bias and mitigated to avoid bias and discrimination.

\item \textbf{Societal and environmental well-being:}
Broader societies, both human and other sentient beings, should be considered in the design of AI systems. This TAI principle of trustworthy AI advises that AI systems should be designed to benefit society and the environment. One key aspect of this is to align the AI design with the global sustainable development goals (SDGs) and promote social responsibility. When AI is introduced in the workplace, specifically when it impacts or replaces human roles, the team members should be considered and consulted in designing the AI system.

\item \textbf{Accountability:}
The principle of accountability states that AI systems should be designed in a way that they can be audited and, therefore, held accountable for their decisions and actions. Monitoring should be possible to do on an ongoing basis, as well as being audited so that redress can be conducted when appropriate. In addition to auditability, This TAI principle covers the area of risk management and risk mitigation for AI systems.
 
\end{itemize}
\subsection{EU AI Act}\label{subsec2}
The EU AI Act\cite{eu2024aiact} requires high-risk AI systems to come with detailed explanations of their design and build, including documentation on their metrics and thresholds for accuracy, robustness, as well as potential discriminatory impacts. They must quantify the degrees of accuracy for specific persons or groups, as well as the accuracy in relation to the specific purpose of the AI system. They must also provide a detailed description of the appropriateness of the performance metrics for the specific AI system. These systems must also include documentation showing compliance with other requirements in the Act. These include establishing foreseeable risks to health and safety or other unintended outcomes, and the AI system's cybersecurity capabilities against things such as data poisoning and adversarial attacks. The act also stipulates that appropriate degrees of transparency and compliance must be reached. The act also calls for the inclusion of the ethics guidelines for trustworthy AI\cite{AIHLEG2019}, when designing and using AI systems.

The EU AI Act stipulates the importance of regulatory oversight and the integration of appropriate safeguards. The act stipulates the importance of the development of standards AI systems must adhere to, which would have to include a balanced representation of interests and involve relevant stakeholders. They also propose the development of national or international regulatory sandboxes to facilitate the testing of AI systems. The Act seeks to enhance existing regulations that exist at a sector level, such as legislation for credit institutes or the medical sector, that do not include AI-specific legislation. 

\section{Methodology}\label{sec3}
This section details the methodology of our study, which employed a systematic literature survey technique. The study was conducted in three phases: (i) actively planning, (ii) conducting and reporting the review results, and (iii) exploration of research challenges. The systematic survey described in this paper followed the widely accepted guidelines and process outlined accepted guidelines and process outlined in Pai et al.\cite{Pai2004} and Kitchenham et al.\cite{Kitchenham2007}. The remainder of this section details the research questions, the process for identifying research, and the data extraction process.

\subsection{Research Questions}\label{subsec3}
The following are the identified research questions (RQs) for this review:
\begin{itemize}
    \item RQ1: What are the main initiatives to establish standards and best practices for determining the trustworthiness of an AI system?
    \item RQ2: What metrics or criteria are currently used to evaluate AI systems for trustworthiness?
    \item RQ3: Are there differences in how each principle of Trustworthy AI is currently being evaluated and scored?
    \item RQ4: Are there works focusing on scoring all the TAI principles?
     \item RQ5: What are the issues and challenges that hinder the development of a scoring and assessment system for the trustworthiness of an AI system?
\end{itemize}

\subsection{Identification of Research (Search Strategy)}\label{subsec4}

To conduct this survey, a search string for Google Scholar was designed to capture papers discussing topics in the machine learning, trust and scoring arenas. Figure \ref{SS} shows the string used.

\begin{figure}[!htbp]
\begin{spverbatim}
("Machine Learning" OR "Artificial Intelligence" OR "Supervised Learning" OR "NLP") 
AND 
("Framework" OR "Scoring" OR "Algorithm" OR "Rating") 
AND 
("Bias" OR "Fairness" OR "Trust" OR "Explainability" OR "Ethics" OR "Governance" OR "Transparency"). 
\end{spverbatim}
\caption{Search String}
\label{SS}
\end{figure}

The initial search returned 971 papers selected for review related to the research topic. A set of inclusion and exclusion criteria, shown in Table \ref{tab1}, were defined to enable the selection of papers to include in this study to be carried out in a systematic manner.  Two researchers independently screened titles in line with Kitchenham et al.\cite{Kitchenham2007}. During the subsequent phase, the full text of the selected papers was thoroughly reviewed. Throughout all three stages, any disagreements regarding the inclusion or exclusion of a paper were resolved through discussions until a consensus was reached. The inclusion and exclusion criteria for papers that contributed to the evaluation criteria classification are detailed in table 1. To ensure all literature review papers covering AI trustworthiness were captured, a second search string term, `Review of the literature trustworthy AI`, was used. A further 11 papers were reviewed to enhance the background, but they did not contribute directly to the paper's main findings. An additional 14 documents were added through citation tracing (snowballing) of some of the critical academic, ISO, and EU publications, culminating in 63 publications contributing to the core findings.

\begin{table}[h]
\caption{Inclusion and Exclusion Criteria for Literature Review on Trustworthy AI}\label{tab1}%
\begin{tabular}{@{}p{0.15\linewidth}p{0.35\linewidth}p{0.35\linewidth}@{}}

\toprule
Criteria & Inclusion & Exclusion\\
\midrule
Topic \newline Relevance & Papers included a metric or discussion of metrics incidentally or explicitly intended to measure any of the seven aspects of Trustworthy AI as defined by the EU. & Papers that neither propose nor discuss potential metrics for the 7 TAI principles. \\
\midrule
 Scope \newline of Research & Papers in various AI domains where trustworthiness is relevant. For principles where metrics are not directly available, papers that discuss potential areas to be evaluated which could contribute to future metrics are included. & Papers that do not propose nor discuss potential metrics for the 7 TAI principles. Papers focusing on simple machine learning or rule-based tasks which do not require advanced ethical evaluation.

 \\
\midrule
 Types of \newline Publication & Academic publications (Journals, conferences, book chapters, thesis workshops, preprint archive), white papers, ISO's or European Commission publications that include a clear methodology, empirical data, and robust analysis. Papers across all disciplines. & Opinion pieces, non-official websites, non-English, duplicates.\\
\botrule
\end{tabular}
\end{table}

\subsection{Data Extraction}\label{subsec5}
The papers were manually reviewed by two authors independently. For each one of the 63 returned papers, the following information was compiled: bibliographic data, the contribution towards the domain of TAI evaluation, the TAI evaluation criteria, the use case
and the type of validation used (if applicable). The data was then compared and aligned, with discussions taking place if any inconsistencies were found.

\section{Review Results}\label{sec4}
In this section, we examine the papers acquired through our systematic review. We summarize their bibliographic details, including the number of publications per year and the publication venues. Subsequently, we dive into a more detailed analysis where we explore the paper content, including the classification of evaluation criteria by TAI principle.

\subsection{Preliminary Results}\label{subsec6}
There has been a notable increase in publications in TAI from 2020 onwards as illustrated in Figure \ref{fig:2}. This aligns with heightened interest from regulatory bodies such as the European Commission.

\begin{figure}[htbp]
    \centering
    \includegraphics[width=.6\linewidth]{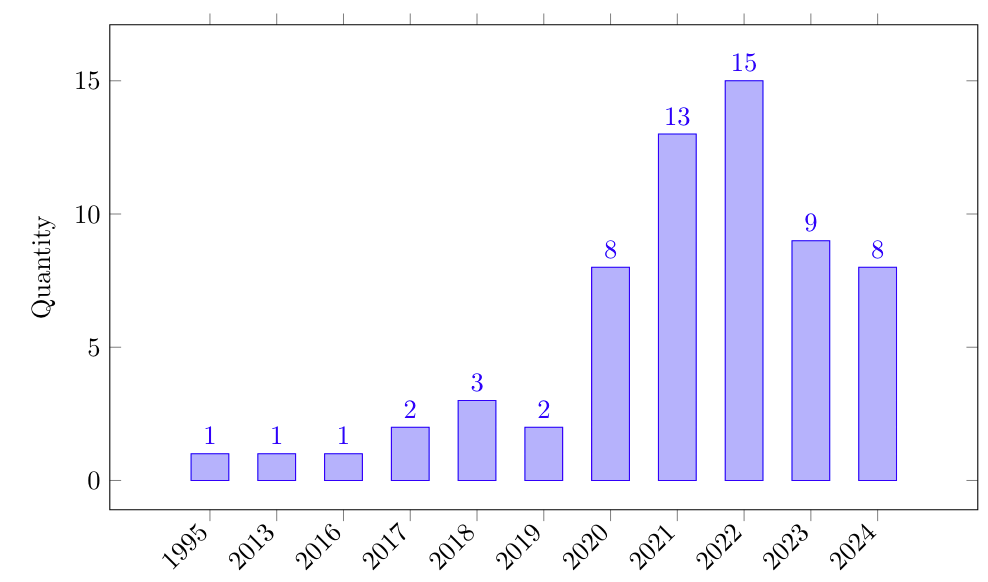}
    \caption{Number of Publications Per Year}
    \label{fig:1}
\end{figure}

The diversity of publication platforms, including several ISO standards and EU TAI publications, reflects the vast array of stakeholders engaged in the field of TAI, spanning regulatory bodies, industry, and academia. This diversity is detailed in Figure \ref{fig:3}.
\begin{figure}[htbp]
    \centering
    \includegraphics[width=.6\linewidth]{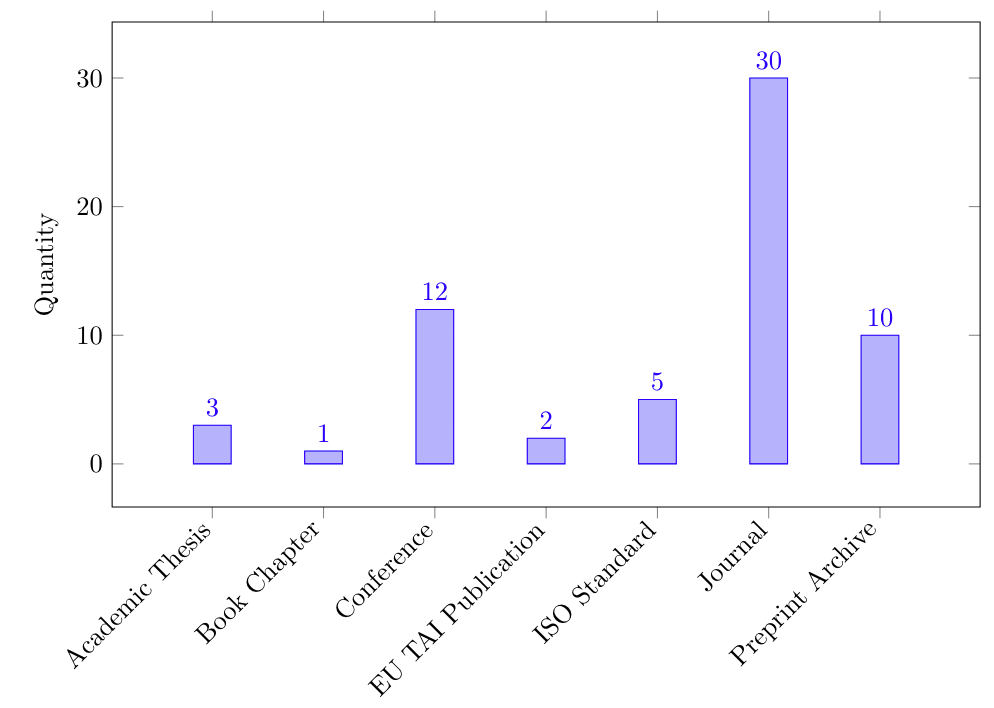}
    \caption{Types of publications}
    \label{fig:2}
\end{figure}

\subsection{Analysis}\label{subsec7}
We identified the papers that included scoring or evaluation of the seven TAI principles. We include a subsection for each principle and a subsection focused on papers that sought to score for multiple TAI principles. Although we include considerations from the more recent and detailed EU documents such as ISO42001\cite{ISO42001} and the AI Act\cite{eu2024aiact} throughout this paper, our primary classification is high-level and informed by the original principles published by the EU, including the EU's ALTAI\cite{EUALTAI_2020}, along with the relevant papers returned from our search. Papers which focused on evaluation methods were only included if they also considered the evaluation criteria or evaluation metrics. There are a number of evaluation methods for TAI which can be classed into automated, semi-automated, manual and conceptual\cite{mccormack2024ethical}. This is outside the scope of this paper, which looks at the evaluation criteria rather than the method used to evaluate it.

\subsection{Evaluation Criteria for Trustworthy AI}\label{subsec8}
\begin{figure}[htbp]
    \centering
    \includegraphics[width=1\linewidth]{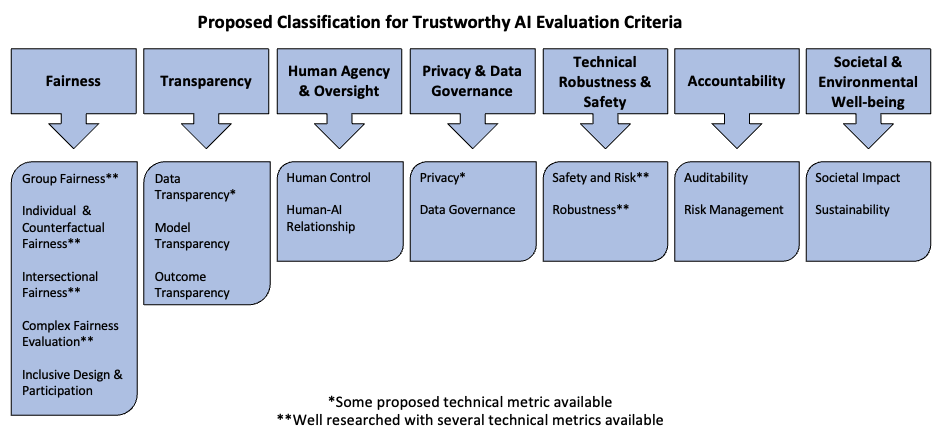}
    \caption{Proposed classification for evaluation criteria for Trustworthy AI}
    \label{fig:3}
\end{figure}

In this section, we use the evaluation criteria in the literature to propose a classification of criteria to evaluate individual TAI principles, including specific metrics where available. We found that most papers returned involved the evaluation of individual principles. In addition to this classification, this section includes a discussion scoring for multiple TAI principles. Our proposed classification is detailed in Figure \ref{fig:3}. Additionally, Table \ref{tab2} seeks to provide further clarity and guidance around key evaluation terminology. These definitions are grounded in the definitions provided in the EU AI Act\cite{eu2024aiact}, with additional context provided in relation to their usage as evaluation criteria based on the findings from this paper.

\begin{table}[htbp]
\caption{Definitions of Terms}\label{tab2}%
\begin{tabular}{@{}p{0.15\linewidth}p{0.95\linewidth}@{}}

\toprule
Term & Definition\\
\midrule
TAI Evaluation \newline Criteria & The considerations for evaluating TAI principles. Metrics are included where available. Alternatively, the most relevant areas for evaluation found in the literature are summarized for each TAI principle.\\
\midrule
Metrics & In this paper, we use the term metrics to refer to quantifiable standards of measurement, typically referring to ones which are calculated mathematically from interactions with an AI system.\\
\midrule
Human Control & An AI system's ability to allow for human intervention, control and ability to stop the AI when necessary, including quality feedback mechanisms. \\
\midrule
Human-AI \newline Relationship & The ability for humans to understand and use an AI system in a satisfactory way, including perceived system trust, user understandability and system satisfaction. \\
\midrule
Safety and Risk & An AI system’s ability to operate without causing harm and ability to be resilient against attacks or attempts to alter the use or performance of an AI system. This includes the system’s ability to mitigate against potential failures in the system. \\
\midrule
Robustness & This covers an AI system's performance, specifically focused on evaluating if it performs consistently and reliably under various conditions, including the presence of unseen data or unexpected scenarios. \\
\midrule
Privacy & The protection for user privacy within an AI system, including the privacy of data used to build the system and that of the end user. \\
\midrule
Data Governance & An AI system's adherence to governance of data collection, data processing, data compliance and data consistency throughout its lifecycle. \\
\midrule
Data Transparency & The capacity of an AI system to facilitate the traceability and explainability of data collection, assumptions, and processing methods. \\
\midrule
Model \newline Transparency & The capacity of an AI system to facilitate the traceability and explainability of the model including model selection, development, and explanation processes. \\
\midrule
Outcome \newline Transparency & The capacity of an AI system to facilitate the traceability and explainability of the outcome of the AI decisions, including the system’s ability to have outcomes challenged. \\
\midrule
Fairness & The term "fairness" is used to refer to the wider area of diversity, non-discrimination, and fairness, specifically looking at an AI’s ability to promote diversity, equal access, and equality, and to avoid discrimination and unfair biases. \\
\midrule
Group Fairness & The effectiveness of an AI system in ensuring fairness across different demographics, in particular protected groups such as gender or race. \\
\midrule
Individual and Counterfactual Fairness Metrics & The effectiveness of an AI system in ensuring individual fairness outcomes and counterfactual scenarios, ensuring fairness for individual users. \\
\midrule
Intersectional \newline Fairness & The effectiveness of an AI system in ensuring fairness in scenarios looking at multiple protected or sensitive user attributes. \\
\midrule
Inclusive Design and Participation & An AI system's inclusivity of design for accessibility for all users, and it’s inclusion of relevant stakeholder participation during the design and operation of the system. \\
\midrule
Societal Impact & How an AI impacts society, including its impact on the workforce, culture, creativity, and potential harm, ensuring positive societal contributions. \\
\midrule
Sustainability & The environmental, social, and economic sustainability of an AI system, including its on-going viability and resource consumption. \\
\midrule
Auditability & How auditable an AI system is, including its available documentation, and the traceability mechanisms in place to ensure accountability for decisions and actions. \\
\midrule
Risk Management & The effectiveness of an AI system in ensuring risk mitigation and that expectations are met throughout the AI lifecycle. \\
\botrule
\end{tabular}
\end{table}

\

\subsubsection{Evaluation Criteria for Diversity, Non-discrimination, and Fairness}
The evaluation of fairness is one of the most researched TAI principles, with many metrics already developed\cite{Paganoetal_2023}. The maturity of this TAI principle is also reflected in its governance. The EU AI Act\cite{eu2024aiact} requires certain AI systems to produce performance metrics, including accuracy metrics for specific people or groups, which is a way to quantify fairness. This section builds on the literature, proposing a high-level classification for the commonly used fairness metrics and several more recently developed fairness metrics. We extracted fairness metrics from the reviewed papers and classed them based on their underlying principles. Additional fairness evaluation considerations, such as accessibility, which are outlined in the ALTAI and have no established evaluation metrics, are also included.
Caton and Haas\cite{CatonHaas_2020} classified metrics into either group fairness or individual and counterfactual fairness metrics. This section expands on this existing classification and proposes three additional high-level classifications: Intersectional Fairness Metrics, Complex Fairness Metrics, and Inclusive Design and Participation Metrics, outlined in Figure 5.

\begin{figure}[htbp]
    \centering
    \includegraphics[width=1\linewidth]{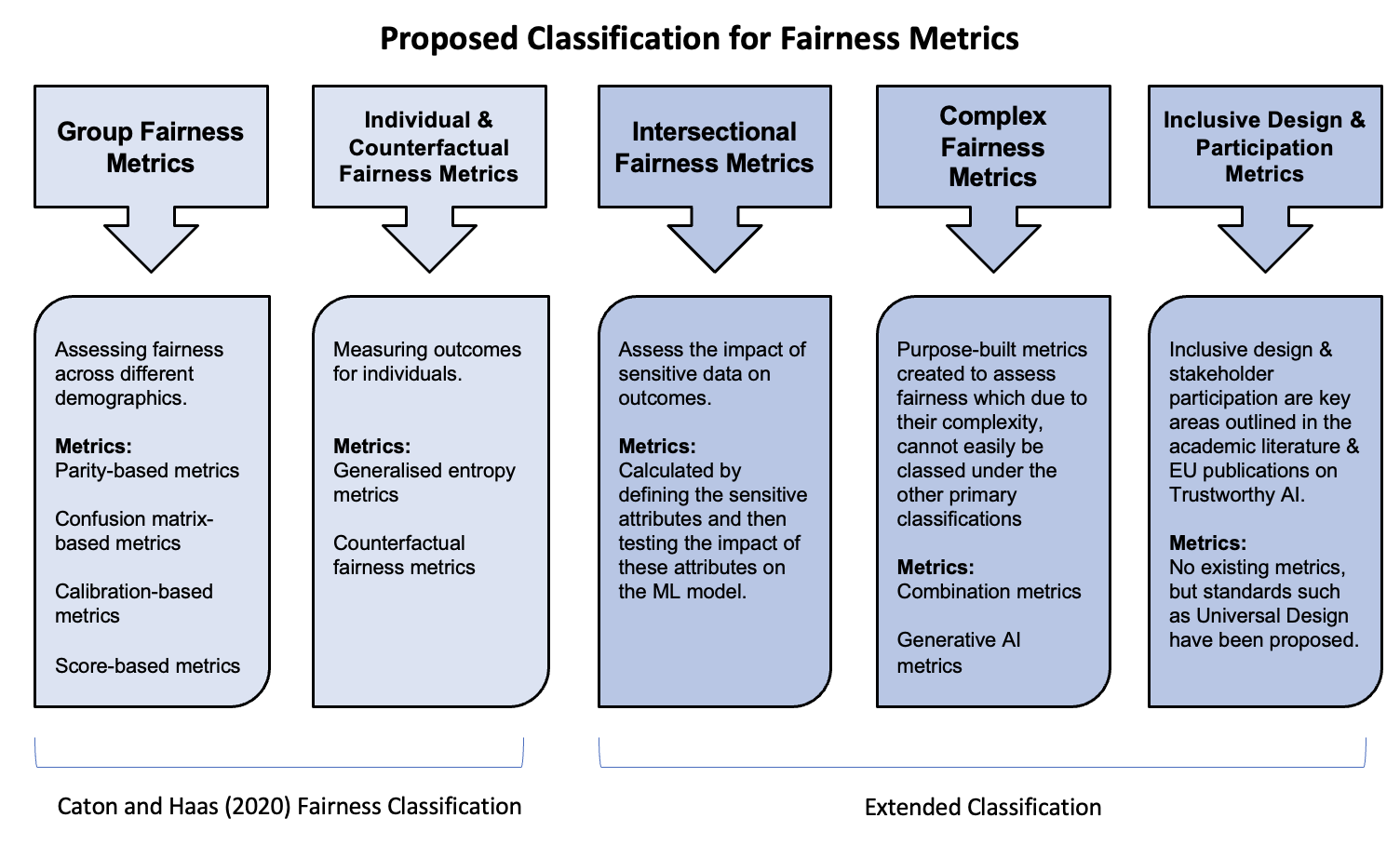}
    \caption{Proposed classification for fairness evaluation criteria}
    \label{fig:4}
\end{figure}
\
\paragraph{Group Fairness and Individual and Counterfactual Fairness }
There are many established metrics for fairness evaluation, with two key papers, Zhou et al.\cite{Zhouetal_2022} and Pagano et al.\cite{Paganoetal_2023}, offering summaries of the usage of several standard fairness metrics. Group fairness metrics included a sub-classification of Parity-based Metrics, Confusion Matrix-based Metrics, Calibration-based Metrics and Score-based Metrics. These metrics all take the standard approach to measure fairness across different demographics, typically for a protected group such as gender or race. Individual and counterfactual fairness methods focus on measuring outcomes for individuals, diverging away from the more traditional group fairness assessments, and include distribution-based metrics like generalized entropy and counterfactual fairness metrics, which can be calculated for individual fairness outcomes\cite{CatonHaas_2020}. 

\paragraph {Intersectional Fairness}
The approaches to fairness evaluation introduced by Zhang et al.\cite{Zhangetal_2020}, Lee et al.\cite{Leeetal_2022}, and Ferry et al.\cite{Ferryetal_2023} provide a case for a new high-level "Intersectional Fairness Metrics" classification. Zhang et al.\cite{Zhangetal_2020} introduced a semi-supervised learning technique to improve fairness, which scored the dependency of a model on a set of protected attributes for their use cases in the areas of health and finance. This combination of techniques does not fit the original classification proposed in Caton and Hass (2020). Similarly, Lee et al.\cite{Leeetal_2022} developed a fairness metric by measuring the dependency or impact of protected classes to assess their technical solution to increase fairness for their use case in crime rate prediction. Their maximal correlation framework sought to improve this fairness metric in ML algorithms by again reducing dependencies between variables, which included protected metrics like race or gender, thus getting a better balance between accuracy and fairness, again showing the need for a new class which involved more complex relationship to multiple pieces of sensitive data. Ferry et al.\cite{Ferryetal_2023} also used a method that primarily focuses on defining multiple sensitive attributes, scoring, and limiting their impact on the model to improve fairness in their robustness framework for statistical fairness in ML. In this paper, the authors used a heuristic application to show how their model can be integrated directly into a range of existing fair classification techniques to improve fairness generalization. Using a sample-robust formulation of the fair learning problem, they aim to meet a given fairness constraint, even when some examples are missing from the datasets used for their use cases around recidivism and finance. The process of assessing fairness by limiting the impact of this sensitive data again makes a case for this new intersectional fairness classification. Maheshwari et al.\cite{maheshwari2024synthetic} proposed an approach to improve fairness in synthetic data generation by leveraging hierarchical group structure, and used intersectional fairness to test their approach.

\paragraph{Inclusive Design and Participation} 
The ALTAI includes the additional fairness considerations of accessible universal design and stakeholder participation\cite{EUALTAI_2020}. Both fairness considerations are also considered in the EU AI Act and ISO/IEC 42001\cite{ISO42001}\cite{EUALTAI_2020}. In the literature, assessing for stakeholder participation is also included in questionnaires that seek to evaluate AI Systems for Trustworthiness\cite{LandersandBehrend_2022}\cite{Chaudhryetal_2022}\cite{Fehretal_2022}. The questionnaire published by Landers and Behrend\cite{LandersandBehrend_2022} also proposed an inclusive design assessment. Due to the inclusion of these fairness considerations in the literature, we suggest including a new classification for fairness evaluation, which assesses both Inclusive Design and Stakeholder Participation.

Although Universal Design does not have established metrics, the ALTAI suggests the ISO standardization for Universal design as an evaluation approach for this metric\cite{EUALTAI_2020}. Landers and Behrend\cite{LandersandBehrend_2022} also proposed evaluating for conformity with the Universal Guidelines and along with other ethical standards.

For stakeholder participation, the ALTAI offers one high-level generic question asking if stakeholders were used in the AI systems design. This question is extended in ISO/IEC 42001\cite{ISO42001}, which asks the organization to ensure relevant experts and human resources are available during the entire AI lifecycle. It also refers to stakeholder considerations such as the requirements of regulatory or other third-party bodies and the impact of the AI system on different stakeholders. Evaluation of stakeholder participation is also included in the latest literature in the area; for example, Chaudery et al.\cite{Chaudhryetal_2022} and Fehr et al.\cite{Fehretal_2022} both propose questions about the inclusion of experts during the AI design and operation stages.  Landers and Behrend\cite{LandersandBehrend_2022} also propose questions about community participation during the design process and the impact of the AI system on stakeholders, including assessing the reaction of those impacted. This area would require further study to develop a more comprehensive quantifiable evaluation approach.

Although there are some technical metrics currently being used to evaluate fairness in ML, there are issues with these, such as a lack of available standardization and benchmarks. Additionally, these metrics focus primarily on assessing ML outcomes for groups or individuals without considering the broader concept of fairness. For a comprehensive fairness evaluation, these wider considerations need to be evaluated. There is a notable difference in maturity for these aspects of fairness which are poorly researched when compared to the evaluation criteria and metrics available for fairness evaluation metrics covered in the existing fairness evaluation classification by Caton and Haas\cite{CatonHaas_2020}.

\paragraph{Complex Fairness Evaluation}
In this section, we include several complex purpose-built metrics created to assess technical solutions proposed to improve fairness. These are primarily created in papers where researchers developed, adapted and combined factors to create fairness metrics to suit their use cases. Due to the complexity of these fairness metrics, they do not fit under the other primary classifications. Thus, we propose a new classification of Complex Fairness Metrics.

Many of these metrics were combined with other metrics, with trade-offs already defined for the specific use case. Lee\cite{Lee_2019} considered commonly utilized fairness metrics such as parity and equalized opportunity flawed. Instead, they measured the aggregate benefit for their use case of loan approval by scoring for increased financial inclusion and measured inequity by scoring the negative impact on minority borrowers. Singh et al.\cite{Singhetal_2021} also developed a custom metric to score fairness in loan approval and expanded on this by providing a benchmark for acceptable fairness. Their universal fairness metric is titled Alternate World Index (AWI) and includes proposed levels for what they considered a balanced trade-off of accuracy and AWI. The proposed standardization of evaluation metrics and the inclusion of suggested levels for acceptable fairness make this fairness metric available for adoption by industry. Khalili et al.\cite{Khalilietal_2021} developed a formula to create a new fairness metric that expands on the equalized opportunity fairness concept, adapting it to include group fairness. Although this was a custom-developed formula to score for fairness, it could be also considered a demographic parity fairness metric. It was tested for multiple use cases, including loan approval and job application fairness. Krasanakis and Papadopoulos\cite{krasanakis2024towards} propose a Phython library called Fairbench, which computes multiple fairness metrics including representation bias, group and individual fairness and parity metrics to create a standardized approach for metrics to assess fairness. 

A number of complex fairness metrics are proposed in the area of generative AI. Due to models being trained on large amounts of uncurated data, there are significant fairness concerns in this area\cite{xiang2024fairness}. Barza\cite{Barza_2023} developed an exciting metric to assess fairness in natural language processing (NLP). They developed a sentence-based evaluation technique to score for fairness using sentence likelihood difference (SLD). The author's NLP bias metric and technique assessed the three classes of gender bias they identified. This process involved creating a Triple Gender Bias dataset, which is used to score for bias with SLD. Platek\cite{platekevaluation} prepared metrics to evaluate Task-oriented Dialogue (ToD) systems and Text-to-Speech Synthesis (TTS). Their open-ended dialog metrics were based on SelfSupervised Learning (SSL) Models, and were enhanced by contrastive losses. When evaluating language, researchers noted complexity due to difficulty in reproducibility\cite{belz2023missing}, along with challenges including pragmatics, semantics and grammar\cite{platekevaluation}. Teo et al.\cite{teo2024measuring} reviewed existing fairness evaluators and proposed a new framework, CLassifier Error-Aware Measurement (CLEAM), which assessed fairness in text-to-image generator and generative adversarial networks (GANs), which includes assessing equal representation and bias. Deldjoo proposed a framework called FairEvalLLM. which evaluated fairness in recommender systems powered by large language models (RecLLMs).  Their framework combines multiple fairness considerations and metrics based on the underlying benefits. These evaluation metrics offer insight into the innovative approaches to tailoring fairness metrics.

\subsubsection{Evaluation Criteria for Transparency}
In contrast to fairness, there is limited research into evaluation criteria for scoring transparency, including a lack of standardization\cite{Alietal_2023}. However, notable categorizations provide a starting point for evaluating AI transparency\cite{Alietal_2023}\cite{Guidottietal_2018}. Bommasani et al.\cite{Bommasanietal_2023} proposed the Foundation Model Transparency Index (FMTI), which includes a comprehensive set of indicators for transparency evaluation relating to multiple TAI principles, including accountability and traceability. Fehr et al.\cite{fehr_2024} also proposed evaluating five key areas: intended use, algorithmic development, ethical considerations, technical validation and quality assessment, and deployment caveats. When considering evaluating transparency, both Bommasani et al. and Fehr et al. proposed scoring the criteria based on public availability and accessibility. However, there are wider considerations needed, such as the variation in transparency requirements by stakeholder, with research\cite{van2020transparency} finding that stakeholders such as engineers and legal experts require different types of transparency.

This paper considers the evaluation of auditability and traceability within the TAI principle of accountability rather than in this transparency section. Additionally, although model performance metrics such as accuracy and reliability are correlated with XAI\cite{Chamolaetal_2023}, these do not evaluate or score the transparency itself; thus, they are also outside the scope of this section.

This section proposes three high-level classifications for criteria to evaluate the transparency of an AI system: Data Transparency, Model Transparency, and Outcome Transparency.

\paragraph{Data Transparency}
Researchers have proposed evaluation considerations around data transparency with a focus on data collection transparency, including transparency of assumptions made, consent and sensitive data collection\cite{Dvoraketal_2021}, \cite{Fehretal_2022}, \cite{Chaudhryetal_2022},\cite{LandersandBehrend_2022}. Data processing transparency, including evaluating if there is disclosure of how data was cleaned and engineered into model features and classes and if any assumption testing was undertaken was also considered\cite{Dvoraketal_2021}, \cite{Fehretal_2022}, \cite{Chaudhryetal_2022},\cite{LandersandBehrend_2022}. ISO/IEC 420001\cite{ISO42001} includes a classification for data in AI Systems. This aims to ensure data transparency to enable the organization to understand the role and impact of data on the AI system. Their four proposed areas for assessment are as follows: data for the development and enhancement of AI systems, acquisition of data, quality of data for AI systems, and data provenance and data preparation. Evaluation of the transparency of data quality throughout the lifecycle enables control over critical steps like data collection, annotation, feature engineering, safeguarding against biases and ensuring the reliability and representation of datasets essential for accurate model training and deployment. Although there are limited metrics available for data transparency, two metrics proposed are calculating the rate of precise data over an entire dataset to get a data accuracy metric and a data diversity metric, which is the ratio between the number of available data sources, their size, and the dataset used\cite{mattioli2024}. Both ISO/IEC 42001\cite{ISO42001} and the questionnaires referenced in this section propose assessing data transparency by checking for documentation and processes relating to the AI system's acquisition, processing and impact of data. The inclusion of these evaluation considerations in the literature and the recent ISO standard highlight the importance of data transparency for AI and the need for this metric in classifying AI transparency.

\paragraph{Model Transparency }
Model behaviour, model selection and model explanation are three key areas for transparency. We included model transparency as a high-level classification for transparency due to the importance placed on three aspects of model transparency in the literature and related ISO standards detailed in this section.

Model behaviour transparency is a critical part of the understandability of an AI system and includes the interrelated areas of model selection, development, and explanation\cite{Alietal_2023}. Model selection, development, and explainability are critical aspects of AI transparency due to their essential role in how AI operates, as highlighted by the literature discussed in this section. This section explains the importance of transparency for model selection, development, and explanation.

Model selection and development is a crucial area for AI transparency as decisions around model selection directly relate to the trade-offs between trust principles, for example, accuracy and interpretability\cite{Lee_2019}. Transparency around model selection and development includes considerations such as appropriateness for intended use, sufficient stakeholder consultation, model selection, and model testing\cite{Fehretal_2022}\cite{Chaudhryetal_2022},\cite{LandersandBehrend_2022}. Model selection should include assessing if alternative architectural designs were evaluated, such as using sampling, experimental design, variable choices, analysis, and interpretation, and assessing if these impacted the validity of conclusions\cite{LandersandBehrend_2022}. ISO/IEC 42001 requires organizations to maintain documentation around model transparency, including the algorithm type, model training, model evaluation and model refinement\cite{ISO42001}. 

Model explanation is another vital part of model transparency and helps people understand how a model works to make decisions or predictions\cite{Bommasanietal_2023}. Although this is a well-researched area, with evaluation criteria for rule-extraction techniques being discussed as early as 1995\cite{Andrewsetal1995}, there is still no standardization or agreement on what a model explanation is for a black box algorithm\cite{Guidottietal_2018}. Despite this lack of standardization, there are many commonly used model explanation techniques, such as SHAP\cite{Lundberg_Lee_2017} and LIME\cite{Ribeiroetal_2016}, albeit with some concerns over their limitations\cite{slack2020fooling}\cite{huang2024failings}\cite{letoffe2024correcting}. ISO/IEC 25059 
\cite{iso25059} suggests using systems documents, logs or introspection tools and data files to improve transparency and states that low transparency systems are ones where internal workings are challenging to inspect externally. 

\paragraph{Outcome Transparency }
The AI legislation, ISO standards and the literature discussed in this section highlight the importance of evaluating the transparency of outcomes of the AI system. This includes transparency around the potential impact on society, outcome explainability, and the ability to challenge the outcome. While there are no metrics developed specifically to quantify AI outcome transparency, this section explains the criteria to be considered for evaluation.

The potentially dangerous impact of AI on society is heavily criticized\cite{Bernsteinetal_2021}. ISO/IEC 42001 aims to help organizations manage their AI systems responsibly and includes impact assessment sub-sections in both the planning and operation sections of the document\cite{ISO42001}. It offers comprehensive considerations for evaluation, including the impact on culture, values, norms, ethics, legal, and contractual or regulatory obligations. It recommends the consideration of these impacts be documented for transparency. The impact of AI systems was also a focus for researchers looking to evaluate AI systems for Trustworthiness\cite{LandersandBehrend_2022}, \cite{Chaudhryetal_2022}. FMTI, the transparency index proposed by Bommasani et al.\cite{Bommasanietal_2023} evaluated the downstream impact of the model on market sectors, individuals, and geographies, as well as future applications. Their results found that downstream impact transparency was virtually non-existent, highlighting this as an area for further research. Both the AI outcome and the ability to challenge the outcome are critical questions for evaluating AI systems for transparency\cite{Dvoraketal_2021}. Commonly used tools like SHAP and LIME can be enhanced through tools like PLENARY, which uses SHAP outputs to explain better the model's decision\cite{Alietal_2023}. Outcome transparency could be scored by checking for the presence of commonly used tools such as those detailed in the taxonomy of interpretability model outputs proposed by Mittelstadt\cite{Mittelstadt_2021} or the explainability methods detailed in Llorca et al.\cite{llorca2024}. Another common way to assess model outcome transparency is through user trust, which researchers frequently use as an assessment criterion to measure the success of introducing improvements in XAI, consistently demonstrating a correlation between enhanced explainability and increased user trust\cite{Druceetal_2021}\cite{Lambertetal_2022},\cite{Guoetal_2022}. This metric is generally assessed by asking the human user of the AI system how much trust they have in the results or decisions provided by the AI system. Jia\cite{Jiaetal_2022} also proposed a method of clinical validation involving an evaluation of user satisfaction, and Feher et al. (2022)\cite{Fehretal_2022} included user trust as a metric in their questionnaire designed to score transparency.  While user trust is correlated, technology, context, social, user-related, and organizational factors contribute to user trust in AI\cite{Yang_Wibowo_2022}.

The transparency assessment frameworks proposed by Bommasani et al.\cite{Bommasanietal_2023} and Fehr et al.\cite{fehr_2024} provide a starting point for evaluating AI transparency; however, they both primarily offer high-level questions to assess the transparency of the system, in conjunction with various aspects of Trustworthy AI. There is a need for more quantifiable metrics, standardization, and benchmarks to assess transparency. A challenge arises due to the varying levels of transparency and explanations required at different stages of AI development and operation, which are demanded by different types of stakeholders\cite{Jiaetal_2022}. Further research would need to be done to develop the existing literature around evaluation criteria and classifications for transparency into a practical scoring mechanism. 

\subsubsection{Evaluation Criteria for Human Agency and Oversight}
 Although specific scoring metrics for this TAI principle have not been heavily researched, criteria to evaluate human agency and oversight as part of TAI evaluations are proposed in the literature. In this section, we propose four key evaluation areas for this TAI principle, which are informed by the ALTAI\cite{EUALTAI_2020}, ISO/IEC 42001\cite{ISO42001}, and by research papers in this area\cite{Fehretal_2022}\cite{Chaudhryetal_2022}\cite{LandersandBehrend_2022}.

\paragraph{Human Control}
The importance of human control is highlighted through the focus given to this area within the ALTAI. The ALTAI proposes both a high-level assessment of societal dependence on a system, as well as individual risks of loss of independence or addiction. Areas to assess for human control include the levels of user control, perceived feedback quality and perceived feedback\cite{Guoetal_2022}). Another consideration for determining human control is establishing if the human has been adequately trained to use the AI system\cite{Chaudhryetal_2022}, ISO/IEC 42001\cite{ISO42001}. The focus on human control in the literature and this ISO standard show the need to evaluate the area of human control. 

An AI system's ability to stop is another element of human control and includes considerations such as whether there are limits identified for where a human is required to intervene and for the AI to stop completely\cite{Chaudhryetal_2022}. There is a requirement for an AI to have the ability to be stopped to limit negative impacts when things go wrong, in particular for autonomous systems impacting humans such as autonomous vehicles\cite{jameel2020ethics}. 

\paragraph{Human-AI Relationship}
The literature highlights that human involvement in AI systems is a crucial area for this TAI principle. Considerations for evaluating the human-AI relationship include evaluating the level of human-in-the-loop involvement\cite{Chaudhryetal_2022}, and the level of human oversight by an expert in the field\cite{Fehretal_2022}. Additionally, training for AI oversight is important for the Human-AI relationship\cite{ISO42001}.

Guo\cite{Guoetal_2022} proposed evaluating user satisfaction, including scoring for perceived system trust and system satisfaction. Evaluating user satisfaction, user trust, and user understandability of the system are considerations for evaluating the human-AI relationship. This overlaps with the area of transparency, which also correlates with user trust (section 4.2.3). The ALTAI also proposes evaluating the TAI principle of Human Agency and Oversight, which overlap with transparency, including assessing the ability of humans to understand the decision made by an AI system and how the system reached its conclusion. We see an additional overlap with transparency in the literature through a positive correlation of user trust metrics with increased transparency and increased human control\cite{Lee_2019}\cite{Guoetal_2022}. Although user satisfaction overlaps with other TAI principles, these points also highlight it as a critical area for assessing the TAI principle of Human Agency and Oversight.

There is limited research on quantifiable metrics for the assessment of the TAI principle of Human Agency and Oversight, however, the literature in this area highlights a need for evaluating specific aspects in the areas of human control, and the human-AI relationship.

\subsubsection{Evaluation Criteria for Privacy and Data Governance }
Privacy and data governance for AI and non-AI systems intersect, so existing policies can be utilized or extended for AI systems\cite{ISO42001}. Evaluating these areas for AI is not a widely researched area. The papers returned from the search string for this paper. For this reason, we did not recommend any change to the high-level classifications proposed in the ALTAI and discussed the evaluation criteria in the literature in this area.

\paragraph{Privacy}
In this section, we include privacy metrics found in the literature that can be used to assess privacy in AI systems.

Differential privacy, as a statistical notion of privacy, aims to provide a way of maximising accuracy while minimising risk and is widely researched and deployed\cite{Khalilietal_2021}. Khalili et al.\cite{Khalilietal_2021} proposed utilising differential privacy to quantify and improve privacy through the introduction of an exponential mechanism which introduces randomness in decision-making. This level of randomness can be controlled through an epsilon ($\epsilon$) score. In this instance, the Epsilon is a parameter used to represent privacy by controlling randomness. A lower $\epsilon$ value means a higher level of privacy due to introducing more randomness, and a higher $\epsilon$ indicates a lower level of privacy and a higher level of accuracy through introducing less randomness. This allows the $\epsilon$ to decide the trade between privacy and accuracy in an ML model.

Lee\cite{Leeetal_2022} and Ferry et al.\cite{Ferryetal_2023} developed metrics to score the impact of sensitive attributes on a model; while their goal was fairness evaluation, this technique could also be used to evaluate privacy. This was undertaken by Li\cite{Lietal_2022} who developed a privacy evaluation and improvement process which was based on evaluating the usage of sensitive data in their model. Other areas which are important for privacy include how data is aggregated, network compression and cryptography\cite{Maetal_2022}), making these areas important for evaluation.

Evaluating the level of data leakage from a model was also researched by Van der Valk and Picek\cite{vanderValkPicek_2019}, and Murakonda et al.\cite{murakonda2020ml} who both looked at assessing the level of data leakage from a model. Van der Valek et at. evaluated how well a malicious agent could learn a secret key from the model. Murakonda et al. built on established algorithms like OpenDP and TensorFlow Privacy, to develop their ML Privacy Meter which quantifies privacy risks to training data for ML models. 

\paragraph{Data Governance}
AI has the potential to significantly change organizational objectives, meaning existing governance may no longer be fit for purpose, and the governing bodies accountable for all activities in an organization may need to change policies to incorporate AI, including data governance (ISO/IEC 38507)\cite{isoiec38507}. This section discusses evaluation criteria discussed in the literature in the areas of data collection, data processing, data compliance and data consistency.

Data collection and processing are key considerations when assessing AI systems for Trustworthiness\cite{Dvoraketal_2021}\cite{Fehretal_2022}, \cite{Chaudhryetal_2022},\cite{LandersandBehrend_2022}. Section 4.3.2 of this paper explained the importance of data transparency for AI. Policies and processes to ensure this type of transparency for data is a crucial area outlined in ISO/IEC 42001\cite{ISO42001} for AI management systems. 

Yan and Zhang\cite{YanandZhang_2022} identified an optimal deterministic algorithm to measure demographic parity, which is helpful for fairness auditing. They proposed future steps to include other metrics within fairness auditing, including equalized odds. In conjunction with the latest fairness metrics detailed in this paper, these steps could be used to develop fairness compliance metrics for privacy and data governance. Shaikh et al.\cite{Shaikhetal_2017} proposed an approach to detect and process natural language policies and flag data usage violations. This conceptual framework involved an AI reading the governance policies themselves and using this to develop metrics to score compliance of an AI system.   

Another evaluation consideration for data governance and compliance is consistency. Hmoud et al.\cite{hmoudetal_2020} assessed the internal consistency of their framework, creating a scoring correlation matrix for each of the items being considered in their framework, which aimed to evaluate the impact and usage of AI in Human Resources systems. To create this scoring framework, they used a process based on Cronbach's alpha to measure internal consistency, combining it with the Fornell and Larckers (1981)\cite{fornell1981evaluating} test to evaluate their framework discriminate validity.  

A significant crossover exists when assessing privacy and data governance for AI and non-AI systems. Further work is needed to extend the existing metrics and standardization to develop AI-specific metrics for privacy \& data governance. 

\subsubsection{Evaluation Criteria for Technical Robustness and Safety}
The TAI principle of technical robustness and safety is widely researched in ML. In this context, we specify evaluation considerations for this principle, including resilience to attack and security, general safety, accuracy, reliability, fall-back plans and reproducibility. 

\paragraph{Safety and risk}
Safety and risk evaluation are well-researched areas, with a number of proposed evaluation metrics in this area. These topics can be considered in line with standard information security evaluation criteria such as those in ISO/IEC 27001\cite{iso27001}.

Simulating security attacks can be used to test the safety capabilities of an AI system. Abramson\cite{Abramsonetal_2020} measured safety against adversarial attacks to assess the success of their proposed ML privacy-preserving trust framework. The authors quantified this by measuring the success of requests made by their malicious agents, which were designed to attempt to create self-signed credentials to request access to their model. Papadopoulos et al.\cite{Papadopoulosetal_2021}, extended this work and noted that this would not be a comprehensive metric for safety against attack, as it doesn't capture attacks coming from authenticated credentials, which may happen in instances where trusted participants became compromised.

Van der Valk and Picek\cite{vanderValkPicek_2019} scored the safety of an AI system by using ML to guess classified information by reviewing how data can be leaked unintentionally. The tool they created evaluated how well the model learned a secret key. The researchers scored and analyse the performance of ML-based Side-Channel-Attacks. By looking at the entropy and bias-variance decomposition, they were able to score and assess what a good or bad result was. Llorca et al.\cite{llorca2024} included an overview of different libraries which can be used to implement different attacks on ML, including a recommendation for public benchmarks to evaluate success. However, they also highlight the risk of developers' gaming systems to achieve the desired metrics.

Privacy metrics such as differential privacy discussed in the previous section are relevant to both the TAI principles of privacy and data governance and technical robustness and safety as this metric can also be used to calculate the security of attacks such as adversarial or inference\cite{Papadopoulosetal_2021}. 

As part of their AI safety evaluation TAI-PRM, Vyhmeister and Castane\cite{vyhmeister2024} introduced a Risk Priority Number index (RPN) for evaluating failure modes' severity, likelihood, and detectability. This metric included a proposed scoring level to determine what is tolerable, treatable or advisable for immediate termination. They also include several other specific risk and safety metrics, including a formula for the Overall Criticality Number (CR), a metric used to assess how critical individual AI assets are to the overall system. These can be used to help evaluate and ensure the safety of an AI system. Ma\cite{Maetal_2022} highlights the additional consideration of computational and communication overheads for security and privacy techniques for AI, outlining its importance for safety evaluation.

Stettinger et al.\cite{stettinger2024} built on the concepts of operational design domain (ODD) and behaviour competency (BC), which are used to quantify residual safety risks in the automated driving domain, to propose evaluating safety boundaries, both during deployment and on an ongoing basis. 

\paragraph{Robustness}
Robustness refers to the ability of an AI system to perform well under variable circumstances, such as using unseen data, and is particularly important given its links to fairness\cite{Adragnaetal_2020}. ISO/IEC TR 24029-1\cite{iso_iec_tr_24029_1_2021} proposes a comprehensive standardized assessment approach for the robustness of neural networks, including a suggested workflow and explanation of evaluation methods. It also proposes three classes of evaluation methods: statistical methods which measure system properties against a target threshold ratio, formal methods which measure performance against sets of fixed metrics, and empirical methods which assess the system in actual or simulated situations. ISO/IEC TR 24029-2\cite{iso_iec_tr_24029_2_2023} includes a further classification of stability, sensitivity, relevance, and reachability, offering detailed evaluation approaches for each. 

Schwarz et al.\cite{schwarz2024} proposed technical robustness metrics, including the Matthews Correlation Coefficient (MCC), Variance Inflation Factor (VIF),  Total Sobol's Variance Ratio (TSVR), and Cosine Similarity Vector Pairs (CSVP). These metrics can be used to assess AI systems' performance, robustness, and reliability across various domains and applications. They also propose the Shapiro-Wilk Test (SWT) and the Breusch-Pagan Test (BPT), which can be used to generate metrics to assess the validity of underlying assumptions of the AI system.

While this area is well-researched for rule-based software systems, further work is needed to develop a comprehensive list of AI-specific evaluation metrics. Traditional software typically operates on static rules. However, AI systems use large data sets and have a more stochastic nature, which introduces complications for validation and operation, requiring a support function similar to a development team\cite{ranjbar2024}.

\subsubsection{Evaluation Criteria for Accountability }
Auditability and risk management are the high-level classifications for accountability proposed in the ALTAI\cite{EUALTAI_2020}. There is currently limited research into evaluating these areas for AI systems. However, given the importance of audibility and risk management in recent AI legislation\cite{eu2024aiact} and standards\cite{ISO42001}, the evaluation considerations that were available in the literature in this area are summarized below using this classification. 

\paragraph{Auditability}
Law and policy focus on transparency, which is intimately connected to accountability\cite{Cooperetal_2022}. Assessing if traceability mechanisms are implemented to improve overall auditability by a third party is a crucial metric for this TAI principle\cite{EUALTAI_2020}. Dvorak et al.\cite{Dvoraketal_2021} proposed assessing AI systems for trustworthiness by checking if the decision-making process step could be identified and separately scrutinized. Although there is a clear need for comprehensive documentation to enhance auditability and transparency, it is unclear how various TAI requirements should be documented together, and the current approach is fragmented with a disconnect between current practices and industry and regulatory requirements\cite{konigstorfer2024comprehensive}. Data provenance is a key aspect of evaluating TAI\cite{ISO42001}, and is important for its auditability. Common aspects such as parameters, algorithm and rationale, the AI's intended use, technical performance evaluation, and limitations are important aspects for auditability\cite{konigstorfer2024comprehensive}. These metrics are included in proposed standardized documentation titled AI Cards proposed by Golpayegani\cite{golpayegani2024ai}. These cards are machine-readable and designed to be transparent and easily auditable. Training data is care aspect for ML models. When considering how auditable an AI system is, we can also consider the TAI evaluation approach of Fehr et al.\cite{Fehretal_2022}, who proposed evaluating if the data used to train a model is available for regulatory bodies.

\paragraph{Risk Management}
To evaluate risk, it is important to evaluate the processes in place to ensure AI training data and outcomes continue to meet the expectations\cite{Dvoraketal_2021}). Assessing if the AI system includes the presence of counterfactuals is another consideration proposed\cite{Chaudhryetal_2022}. 

The ALTAI recommends assessing risk management by looking at whether the company using the AI system has considered the other six areas of trustworthy AI, and specifically, if they have this well documented, including documented processes for the other TAI principles. AI system risk can be evaluated based on the quality of relevant data provenance. ML development should include a datasheet during pre-processing, a Models Evaluation Report, and a disclaimer document around the model's usage context\cite{Chaudhryetal_2022}. ISO 42001\cite{ISO42001} includes a requirement for staff to have the ability to report concerns both confidentially and anonymously, during the entire lifecycle of the AI process, referring to ISO 37002\cite{iso37002} for further clarity on this expectation.

While the ALTAI highlights both audibility and risk management, there has been minimal research into developing a metric to score for this TAI principle.

\subsubsection{Evaluation Criteria for Societal and Environmental Well-being}
Due to the lack of maturity in this field, we have proposed our classification based on the key areas suggested for evaluation in the literature, societal impact and sustainability. Although these considerations are discussed in the literature, to the best of our knowledge, there are no previous metrics developed in this area.

\paragraph{Societal Impact}
In this section, we discuss four high-level areas for assessment, which we found in the literature: workforce impact, culture and creativity impact, knowledge sharing, and potential for harm.

A key concern in this area in the literature is the impact on the workforce. There are concerns about AI systems' lack of humanity and human feelings and what this means as we increasingly replace human jobs with AI\cite{jameel2020ethics}). Köse (2018)\cite{kose_2018}discussed ensuring AI is only used for jobs it is deemed suitable for, with an additional concern around AI systems relying on underpaid workforces. Considerations like job displacement and fair labour are criteria to consider when evaluating AI systems for their impact on the workforce.

A second area for evaluation is the impact on culture and creativity. AI will impact the transformation of human social lives, introducing a risk of a loss of culture and conflicts of ownership in creative spaces such as the production of art\cite{kose_2018}. Landers and Behrend\cite{LandersandBehrend_2022} propose assessing if AI systems have considered the broader cultural context, though this question has not been developed further. We can see variations in approaches to AI governance based on the differences in societal values\cite{Kozuka_2019}. While it may be challenging, the literature does highlight a need to evaluate the impact of AI on culture and creativity. 

Feher et al.\cite{Fehretal_2022} suggested assessing AI systems for their availability of training data to researchers. The ability of an AI system to share its data, insights and algorithms with researchers is a consideration for evaluating the social benefits of an AI system.

AI systems can be used for a harmful purpose through humans selectively applying AI, undercutting their validity, or using ethical systems such as facial recognition for unethical reasons\cite{Bernsteinetal_2021}. There is a need for the accountability of personalization algorithms, which are used to deliver tailored news content, user-generated content, and advertisements to individuals\cite{Descampeetal_2022}. An AI system's ability to facilitate disinformation is a threat to democratic values, which could lead to considerable negative effects on an entire city, domain activity or community\cite{eu2024aiact}. This is another important evaluation criterion for measuring the potential for harm to society.

\paragraph{Sustainability}
In this section, we combine multiple AI sustainability considerations highlighted in the literature, including social, environmental, and economic sustainability. Evaluating sustainable AI can be based on its ability to adapt and exist over a long period by being built to anticipate needs and mitigate the risks associated with societal well-being\cite{deAlmeidaetal_2021}. In addition to this social sustainability evaluation approach, we found researchers focusing on the environmental perspective of sustainability. Environmental considerations include computational overheads, AI systems powered by renewal energy, and AI resource consumption\cite{Hauptetal_2021},\cite{Chamolaetal_2023},\cite{Maetal_2022}. High-level economic sustainability considerations relating to AI were highlighted by Pagano et al.\cite{Paganoetal_2023} and Chamola et al.\cite{Chamolaetal_2023}. In order to class AI systems, the EU AI Act\cite{eu2024aiact} suggests measuring the level of mathematical operations, known as floating point operations (FLOPs), to calculate the amount of computation used for training the model as an alternative to measuring estimated energy consumption for the training of models. Although the act advises that AI systems should seek to adhere to the seven ethical AI principles, including environmental well-being, as well as highlighting that a high level of environmental protection is a fundamental right, it does not stipulate any detailed evaluation criteria for AI systems relating to environmental well-being. 

While this TAI principle has crossover with other areas, such as fairness, evaluating an AI system for its impact on societal and environmental well-being is not well-researched and requires more work to develop evaluation criteria for the proposed classification. 

\subsection{Assessing for Multiple Trustworthy AI Principles}\label{subsec9}
While most of the research looked at evaluating single principles of AI trustworthiness in isolation, several papers looked at a combination of TAI principles. This is important when you consider how the various TAI principles are linked. Existing research into improving TAI showed several principles that were frequently researched together. Diversity, Non-discrimination, and Fairness were among the most researched areas. It was often considered in conjunction with improving transparency, human agency and oversight, privacy, and accountability. 

Agarwal (2020)\cite{Agarwal_2020} proved that principles like fairness, transparency, privacy, and robustness must be considered simultaneously because they impact each other. The author found that a trade-off must be made between the various principles tested for their use case.     

Fairness and robustness metrics in ML are often considered in conjunction with each other\cite{Adragnaetal_2020}\cite{Leeetal_2021}\cite{Liuetal_2022}\cite{Ferryetal_2023}. As well as measuring the impact of the sensitive attributes on the model, both Ferry et al.\cite{Ferryetal_2023} and Lee et al.\cite{Leeetal_2022} quantified the robustness of ML models as part of their assessments on their proposed methods to improve fairness, finding a correlation between the two. Adragna et al. (2020)\cite{Adragnaetal_2020} also found a mutually beneficial relationship between robustness and fairness for their use case. 
  
Lee (2019)\cite{Lee_2019} proposed context-conscious fairness, which considers the trade-off between accuracy and interpretability and the trade-off between aggregate benefit and inequity. Their solution involved benchmarking the trade-offs outlined to make context-based informed choices. They aimed to create transparent decision-making considering the areas of bias and fairness for their use case of mortgage approval algorithms.  Lee and Floridi (2021)\cite{LeeandFloridi_2021} looked at developing use case-specific metrics where decision-makers choose the trade-off between competing objectives when making decisions for their use case, mortgage lending. They scored five algorithms using the metric of equality of opportunity, along with five other metrics relating to fairness and accuracy. They aimed to create a use case-specific algorithm that outperformed other models on multiple dimensions. Their process required humans to decide on a trade-off for their use case. These trade-offs were mapped and could be used for things like accountability. 
  
Yanisky-Ravid and Hallisey (2018)\cite{YaniskyRavidHallisey2018} proposed that increased transparency would also result in increased privacy and fairness. Although their framework was theoretical, they proposed a transparency model centred on the disclosure of data being used by the AI system to improve both metrics rather than needing a trade-off.  
  
Whether optimising models for multiple metrics is mutually beneficial or requires a trade-off, it is undoubted that there will be some cost to business to optimising models for trustworthy metrics. While further study is needed on the relationship between trust metrics and business goals, Kozodoi et al. (2022)\cite{kozodoietal_2022} developed a process to look specifically at the trade-off of fairness and profit in credit scoring. They developed fairness and profit metrics for their use case and showed that fairness could be achieved relatively cheaply to profit for this use case.

\subsection{Case Study: Mortgage Lending}\label{subsec10}
In this section, we look at the selection of evaluation criteria for multiple TAI principles for the use case of credit lending. This use-case is well researched for fairness evaluation. Lee and Floridi\cite{LeeandFloridi_2021} detail their selection process for fairness evaluation metrics, ultimately proving that there were no existing fairness metrics suitable for their use case. While the authors detail the flaws in commonly used parity-based metrics such as equalized odds, which assess for group fairness across demographics, they were also able to utilize existing metrics to develop a new metric which addresses those limitations. As detailed in section 4.3.1, complex fairness metrics such as those proposed by Lee and Floridi have also been built by other researchers in this area. The laws surrounding mortgage lending play an important role in shaping the selection of appropriate metrics for this use case, however each use case presents its own complexities. The classification in \ref{fig:4} provides a starting point for those looking to expand on evaluating not just fairness, but also the other principles of trustworthy AI. This figure also highlights which evaluation criteria have existing metrics available for consideration. For the use case of mortgage lending, we can consider combining established fairness metrics, with other established metrics, such as technical robustness and safety metrics. While these metrics could be combined in an automated process where trade-offs can be established, automation will not be possible for evaluation where no metrics exist. For the use case of mortgage lending, the remaining principles can be evaluated using the proposed criteria in this paper, with high level considerations outlined in table\ref{tab3}.

\FloatBarrier

\begin{table}[htbp]
\caption{Proposed TAI Evaluation Considerations in Mortgage Lending}\label{tab3}
\begin{tabular}{@{}p{0.2\linewidth}p{0.7\linewidth}@{}}

\toprule
Evaluation Criteria & Considerations\\
\midrule
Fairness & While existing complex fairness metrics\cite{Lee_2019}\cite{Singhetal_2021} exist for credit lending, these need to be expanded to consider things such as accessibility or Universal Design principles. \\
\midrule
Transparency & An AI system for mortgage lending should be scored for its data transparency (data collection from mortgage applicants using the model, as well as data used to train the model from previous applicants, assumptions, consent \& processing), model transparency (selection, development, accuracy, explainability) and outcome transparency (model outcome, interpretability, and ability to be challenged where needed, transparency of impact on broader society). \\
\midrule
Human Agency and Oversight & Evaluating societal dependence of an AI based mortgage lending system, evaluating to what extent the system facilitates human control (overriding or challenging AI decisions, or stop the system when needed), evaluating the human satisfaction with the system, both from the banking level and the mortgage applicant level. \\
\midrule
Privacy & Metrics based on the concept of differential privacy\cite{Khalilietal_2021} exist which could be adapted for this use case to establish trade-offs between other existing metrics like fairness and robustness. \\
\midrule
Data Governance & Evaluation of the data provenance quality (data collection, assumptions, transformation) and compliance with relevant laws and standards\cite{eu2024aiact}\cite{ISO42001}. Automated approaches to policy evaluation are proposed in the literature\cite{Shaikhetal_2017} which could be developed for this use case. \\
\midrule
Safety and Risk & Risk Priority Number index (RPN) can be used to evaluate failure modes' severity, likelihood, and detectability\cite{vyhmeister2024}. Additional metrics exist which could be adapted to test the AI system's ability when faced with adversarial attacks\cite{Abramsonetal_2020}\cite{Papadopoulosetal_2021}\cite{vanderValkPicek_2019}. \\
\midrule
Robustness & Depending on the AI system being used, model performance under various conditions can be evaluated using metrics such as Matthews Correlation Coefficient (MCC), Variance Inflation Factor (VIF), Total Sobol's Variance Ratio (TSVR), and Cosine Similarity Vector Pairs (CSVP) proposed by Schwarz et al.\cite{schwarz2024}. \\
\midrule
Accountability & The AI system should be evaluated for its ability to be audited, and the risk management processes in place which should be in line with the relevant standards\cite{eu2024aiact}\cite{ISO42001}.\\
\midrule
Societal Impact & The AI system should be evaluated for its impact on the workforce, culture and creativity, its ability to share knowledge for good (such as sharing mortgage data with researchers or law makers), and its potential for harm. \\
\midrule
Sustainability & Evaluation of social, environmental, and economic sustainability should be done on the AI system. This includes anticipating risks, evaluating its ability to adapt over a long period of time, and evaluating its environmental impact. Computational resource requirement can be calculated using FLOPs\\
\botrule
\end{tabular}
\end{table}
\FloatBarrier

\section{Issues and Challenges of Scoring Trustworthy AI}\label{sec5}
There has been an increase in publications in Trustworthy AI in the past five years, with many active stakeholders in this area. The foundational work was laid in 2018 by the non-profit research group AI4People, which published its ethical AI framework\cite{Floridietal_2018}. This framework led to joint initiatives by researchers and the European Commission, which published the initial guidelines for Trustworthy AI in 2019. The first draft of the AI Act has led to the development of ISO Standards in this area, along with industry white papers on this topic. The complexity and number of stakeholders in the area of TAI have led to a number of challenges, including variance in the maturity of research for each principle and a shortage of appropriate evaluation criteria and metrics. There is a gap in evaluation to assess multiple TAI principles. There is no consistent approach to evaluation, with little standardization of TAI evaluation criteria, metrics and benchmarks. This is partially due to the requirement for tailored approaches to the evaluation of the Trustworthiness of AI systems per use case. Within organizations, there is a need for technical, ethical expertise in companies where TAI self-assessment is required.

\paragraph{Research Lacks Maturity in Scoring TAI Principles}
Some principles are more researched with a number of technical evaluation metrics available to assess two principles of fairness and technical robustness and safety, and a handful of metrics available for the two TAI principles of privacy and ata Governance and Transparency. The principles of human agency and oversight, accountability and societal and environmental well-being are the least mature in terms of evaluation criteria, with no metrics available in this area. 

Fairness stands out as the most researched principle, with a large number of existing metrics available. There are also proposed benchmarks and metrics for fairness for specific use cases, such as loan approval. However, even within this more developed area, there is a lack of research considering a complete fairness assessment, focusing primarily on technical assessments of ML models used for predictive outcomes such as recidivism, healthcare, and finance. Technical robustness and safety, as well as privacy and data governance, also have technical metrics available but are not as well developed as fairness. Assessment for the TAI principle of societal and environmental well-being is not well researched. Despite the need to measure things like impact on society and sustainability, the work in this area extends only to discussions around evaluation criteria and what might be important to consider. There are no approaches that extend to developing metrics to quantify this. The same is true for evaluating accountability, and human agency and oversight. The Foundational Model Transparency Index (FMTI) detailed in section 4.3.2 includes a transparency scoring index, with a published evaluation of the transparency levels of prominent AI systems like those of Amazon and OpenAI. While there are no automated metrics that can be calculated from interactions with an AI system for transparency, FMIT offers a starting point for evaluation criteria for transparency.

\paragraph{Lack of Appropriate Metrics}
Across all principles where metrics were being used, researchers typically developed those metrics to measure the effectiveness of their work across TAI principles. However, there is a lack of focus on papers looking specifically at identifying appropriate metrics. In many cases, researchers did not explain their choice of metric or failed to demonstrate consideration of alternative metrics. This is important considering the EU AI requires certain high-risk AI systems to detail documentation of their choice of performance metrics, including their suitability for use in that particular AI system. In the more mature field of fairness evaluation, researchers detailed why certain metrics were unsuitable for a specific use case and developed their own metrics which were better suited. Given that the AI Act requires certain high-risk AI systems to select and justify the use of metrics to quantify things like overall accuracy, accuracy for specific persons or groups, and robustness, the lack of research into the selection of evaluation metrics presents a significant gap in the literature.

\paragraph{Lack of Metrics Addressing Multiple TAI Principles}
While some papers measure multiple principles, most focus on individual TAI principles, separating research by principle is a significant concern. However, a key finding in this paper is that principles are interrelated. While this sometimes results in a positive correlation, in many cases, a trade-off between principles is required. Positive correlations found by researchers include; increasing the robustness of a system tends to improve fairness outcomes, and increasing transparency tends to improve areas of human agency and oversight. However, other researchers found that principles were negatively correlated, and trade-offs must be made between TAI principles and business. Under the EU AI Act, documentation must be provided for decision-making processes for both metric selection and trade-offs relating to compliance with legislation. It is at the core of business to make decisions around prioritization and trade-offs based on the goals and values of a business. If that business is using AI as part of those trade-offs, in many cases, those decisions can and must now be documented. The most commonly researched TAI principle is fairness, which typically correlates with accuracy. As you retrain a model to improve fairness metrics such as group parity for protected groups, this typically impacts the accuracy of the model negatively. When you add in metrics for other TAI principles, such as using differential privacy to evaluate data privacy or scoring the success of adversarial attacks to contribute to an evaluation of system safety, the trade-off decision becomes more complicated. For example, if you calculated the computational resource required from an AI system using FLOPs, and it showed one model resulted in a significant increase in usage of computational resources, but it made the system slightly more accurate, fairer or improved the data privacy or security of the system, a decision would have to be made around the environmental impact, versus those metrics. The lack of metrics available for certain principles is a problem for trustworthy AI, because these performance metrics are inherently linked to things like model selection and how the model is trained. Additional TAI evaluation metrics need to be developed so that decisions can be made around the ethical design and development of AI systems.

\paragraph{Necessity for Tailored Approaches and Metrics} Researchers found that different approaches, metrics and thresholds for TAI principles are required depending on the application or use case. The results of TAI evaluation can influence the system design, making it inherently difficult to design an evaluation criteria suitable for multiple use cases. While the seven principles provide a practical high-level framework when assessing AI systems for trustworthiness, the selection of specific appropriate metrics is required for individual use cases. The EU AI Act highlights the importance of the development of standards and notes the overlap with existing legislation at a sector level. It gives the example of regulation for medical devices or credit lending institutes, which have existing regulations which are not AI-specific. Given the finding that each AI use case requires its own evaluation criteria, it is important that sector-level regulator bodies engage in the development of appropriate AI evaluation criteria. This should include the development of use case-specific metrics and acceptance levels for thresholds and trade-offs between these. For example, the most suitable metrics to evaluate fairness, robustness, accuracy and data privacy, and the acceptable thresholds for trade-offs between those will not be the same for both a medical device and a credit lending algorithm. It is not sufficient for AI evaluation criteria and metrics to be developed in a one-size-fits-all model; each sector will be required to develop its own TAI evaluation criteria for each of its various use cases.

\paragraph{Need for Technical Ethical Expertise and Alterations to Organizational Structure} 
The governance of existing IT and data management infrastructures is designed based on static systems that typically use rule-based programming. AI systems are more stochastic and use large datasets to generate dynamic outputs. Organizations seeking to enable effective incident reporting and response management must adapt or establish department functions trained in ethical AI. These teams would be required to be highly agile and responsive, capable of swiftly addressing emerging ethical issues, effectively operating more like a development team than a typical compliance department\cite{ranjbar2024}. Over half of organizations have adopted AI in multiple departments. AI adoption is continuing to rise significantly, with a noted spike in the adoption of generative AI, which doubled in a period of just 10 months from 2023 to 2024\cite{McKinsey2024}. This swift uptake of AI use comes with changes in how businesses are structured and the need for appropriate governance and risk mitigation strategies. Unlike the EU legislation on data privacy, which requires certain organizations to create roles such as a data protection officer\cite{lambert2016data}, the AI Act\cite{eu2024aiact} does not include any legal requirement for the establishment of core teams or roles for TAI. Although non-mandated by the AI Act, there is a clear need for significant changes to organizational structure to ensure AI systems are designed, built and operated to meet the requirements of the act, including its recommendation to adhere to the seven principles for TAI.

\paragraph{Need for Standardization} Both the AI Act and the literature show a need for the standardization of metrics to evaluate TAI principles, with a particular focus on creating acceptable performance metrics for fairness, accuracy, safety and robustness. Existing metrics exist for each of these, meaning that mathematically quantifiable metrics can be calculated from direct interaction with the AI system, forming tangible, reproducible, standardized metrics. However, these metrics would likely impact each other, requiring standardization of metrics to include not only acceptable metrics and thresholds but also trade-offs between them. Additionally, there are no performance metrics created for many of the TAI principles, meaning that the standardization process for these can not include performance metrics produced from interaction with the AI system. While the AI Act does mandate high-risk AI systems to produce a number of quantifiable metrics, including fairness metrics like the accuracy of the model across groups, most of the standards it requires will be non-metric-based assessment standards such as ISO/IEC42001. Where TAI evaluation metrics, which are calculatable through interaction with the AI system are not available, the evaluation criteria must be decided and subsequently evaluated in a more manual way, such as through ISO certification. This is a problem as evaluations using these types of manually assessed standardisation are not fit to evaluate the complexities and unknown risks that AI systems present. Therefore creating universally acceptable standardized metrics for Trustworthy AI that are able to be easily calculated and monitored is critical for the benchmarking, compliance, and governance of AI systems. Additionally, societies have different beliefs on what is fair and ethical, resulting in variances in AI governance approaches, offering challenges to setting universally accepted standards for trustworthiness. Additionally, each use case requires its own standardization of metrics to be developed. In order to truly ensure AI systems comply with the seven TAI principles, significant work is needed to develop standardized metrics which are interrelated and mathematically calculable through interaction with the AI system.    

\section{Conclusion and Future Directions}\label{sec6}
In this paper, we utilized an established systematic review methodology to summarize and classify metrics for evaluating the trustworthiness of AI systems. There is a pressing need for a systematic, multi-dimensional approach to establish TAI evaluation metrics calculated through interaction with an AI system, building on the existing frameworks proposed by the EU AI Act, ALTAI, and relevant ISO/IEC standards. While research on evaluation criteria for the TAI principles is often conducted for individual principles in isolation, the literature also demonstrates that these principles are interrelated and frequently exhibit positive or negative correlations, necessitating trade-offs. As AI systems evolve and become more complex with advanced technologies and additional layers of stakeholders, the challenges of evaluation grow. Moreover, different use cases require specific metrics, thresholds, and approaches for Trustworthy AI evaluation. Given AI's significant societal impact, there is a clear need for consensus on metrics and benchmarks for Trustworthy AI tailored to individual use cases, regulated at a sector level. This paper lays a foundation for more rigorous, standardized, and context-sensitive evaluations and scoring of AI trustworthiness, with far-reaching implications for research, policy, industry, and society.

\backmatter



\bmhead{Conflict of Interest}
On behalf of all authors, the corresponding author states that there is no conflict of interest.

\bibliography{main}

\end{document}